\documentstyle{mn} 
\input epsf.tex

\begin{document} 

\title[On the spin paradigm and the radio dichotomy of quasars]
{On the spin paradigm and the radio dichotomy of quasars}

\author[Moderski et al.]  {R.~Moderski~$^{1,2}$, M.~Sikora~$^2$
and J.-P.~Lasota~$^1$\\
$^1$ UPR 176 du CNRS; DARC, Observatoire de Paris, Section de Meudon,
92195 Meudon, FRANCE\\
$^2$ Nicolaus Copernicus Astronomical Center, Bartycka 18, 00-716 Warsaw,
POLAND} 
\date{Accepted . Received ; in original form .} 
\volume{000}
\pagerange{000--000} 
\pubyear{1998} 
\maketitle 
\begin{abstract}
  We investigate whether models based on the assumption that jets in
  quasars are powered by rotating black holes can explain the observed
  radio dichotomy of quasars. We show that in terms of the ``spin
  paradigm'' models, radio-loud quasars could be objects in which the
  black hole's rotation rate corresponds to an equilibrium between
  spin-up by accretion and spin-down by the Blandford-Znajek
  mechanism. Radio-quiet quasars could be hosting black holes with an
  average spin much smaller than the equilibrium one. We discuss
  possible accretion scenarios which can lead to such a bimodal
  distribution of black hole spins.
\end{abstract} 
\begin{keywords} accretion, accretion discs --
black hole physics -- galaxies: active -- galaxies: evolution.
\end{keywords} 
\section{INTRODUCTION} 
Quasars are characterized not only by an intense radiation from the
central engine, but also by jets which power large scale radio
structures. The ratio of radio luminosity of these structures to the
optical luminosity of the central sources shows a bimodal
distribution, with only $\sim 10$ \% of quasars belonging to the
radio-loud category (see, e.g., Kellerman et al. 1989; Hooper et
al. 1995; Falcke, Sherwood, \& Patnaik 1996; Bischof \& Becker 1997).
Radio-loud quasars have never been found in spiral galaxies, while the
hosts of radio-quiet quasars are either spiral or elliptical galaxies
(Taylor et al. 1996; Kukula et al. 1998).  On the other hand, both
radio-quiet and radio-loud quasars have almost identical average
IR-optical-UV spectra (Francis et al. 1993; Zheng et al. 1997), which
suggests similar accretion conditions in these two samples of objects.

These properties seem to support models which are based on the idea
that jets are powered by rotating black holes.  However, in order to
explain the observed radio loudness bimodality in terms of a bimodal
distribution of black hole spins, one has to assume that the
population of supermassive black holes is dominated by very low-spin
black holes, and one should understand why radio-loud quasars are
always hosted by elliptical galaxies.  This issue has been addressed
by Wilson \& Colbert \shortcite{wc95}, who proposed that high-spin
black holes exist only if formed by coalescence of two black
holes. Such coalescences would take place mostly in a denser galactic
environment (groups, clusters), which are much more populated by
elliptical galaxies than the field regions.  However, the basic
assumption of this scenario, that black holes which do not undergo
coalescence rotate very slowly -- despite the angular momentum gained
from accretion discs -- must be verified. In other words, one should
verify that there exists a mechanism which could keep black holes at
low rotation rates despite of the accreted angular momentum.

This problem was investigated by Moderski \& Sikora
\shortcite{ms96a}. They showed that the Blandford-Znajek (B-Z)
mechanism, which extracts rotational energy from a black hole
\cite{bz77}, is not efficient enough to counteract the spinning up of
a black hole by the gas accreted from standard $\alpha$-discs
\cite{ss73}.  Low-spin equilibrium states are possible only for very
low accretion rates but the time required to established such an
equilibrium is longer than the age of the Universe.  This is because
the power extracted from the black hole is proportional to the square
of intensity of the magnetic field which threads the black hole.  The
field intensity is limited by the pressure in the accretion flow and,
therefore, is very low for very low accretion rates. Hence, once a
black hole gets a high spin, it will be rotating fast forever.

Low-spin equilibrium solutions for high accretion rates are possible
only for the so called $\beta$-discs \cite{ms97}.  In such discs the
viscous stress driving accretion is proportional to the gas pressure
only (and not to the total pressure as in the case of $\alpha$ discs).
Since the relation between the viscous stress and the pressure in
radiation pressure dominated accretion discs is unknown, such discs
are a reasonable alternative to $\alpha$-discs. For high accretion
rates the gas pressure is by several orders of magnitude smaller than
the radiation pressure \cite{ss73}.  This implies that, for a given
accretion rate, $\beta$-discs are much denser than $\alpha$-discs
\cite{sc81} and, therefore, can confine much stronger black hole
magnetic fields.

Since the power extracted from a black hole via the B-Z mechanism
scales with the square of the magnetic field intensity, equilibrium
spins for $\beta$-discs are smaller than for $\alpha$-discs. This, of
course, does not mean that the power extracted from a black hole with
a $\beta$-disc must be smaller than the power extracted from a black
hole with an $\alpha$-disc, because in the expression for the B-Z
power a lower equilibrium spin is compensated by a higher intensity of
the magnetic field \cite{ms97}. Thus, low equilibrium spin black holes
can represent radio-quiet quasars only, if the conversion of the
extracted black hole energy into jet energy is very
inefficient. According to the Wilson-Colbert scenario powerful jets
would then exist only in objects where a coalescence of two black
holes leads to spins much higher than the equilibrium one. However,
powers extracted from black holes with spins much larger than the
equilibrium one are so large, that such black holes would be spun-down
on time scales two orders of magnitude shorter than the typical
lifetime, $\sim 10^8$ years, of radio quasars (see, e.g., Leahy,
Muxlow \& Stephens 1989).

Powering jets in radio-loud quasars by a black hole can last as long
as $10^8$ years, only if losses of angular momentum due to the B-Z
mechanism are compensated by gains of angular momentum from an
accretion disc. This would give a correlation between radio and
optical luminosities, in accordance with observations
\cite{ser97}. Then, however, one would have to explain why the
majority of super-massive black holes have spin values much lower than
the equilibrium one -- the condition for the existence of radio-quiet
quasars. As suggested by Moderski, Sikora \& Lasota \shortcite{msl97},
black holes in most objects could be forced to rotate slowly by
multi-accretion events with random orientations of the angular
momentum vector. In such a scenario, quasars which become radio-loud
are only those which undergo major accretion events, induced, i.e., by
a merger of two big galaxies.  Following this, a black hole can easily
double its mass and reach an equilibrium state.  In this paper we
explore this possibility and derive conditions the model must satisfy
in order to explain the observed radio properties of quasars.

This paper is organized as follows. In Section 2 we present
equilibrium spin solutions, obtained for a variety of accretion disc
models. In Section 3, we discuss multi-event accretion scenarios. In
Section 4 we use our results to derive conditions which must be
satisfied by quasar evolution models in order to obtain bimodal
distribution of black hole spins.
\section{EQUILIBRIUM STATES} 
The evolution of a black hole, as determined by a gain of energy and
angular momentum from an accretion disc and a loss of energy and
angular momentum via the B-Z mechanism, is described by equations
\cite{ms96a}
\begin{equation}
{dA \over dt} = (\bar j_{in} - 2 A \bar e_{in}) {\dot {\cal M} \over
M} - \left ( {2 \bar r_h \over  k A} - 2 A \right ) {P \over M c^2},
\label{evolA} 
\end{equation} 
\begin{equation} {d \ln M \over dt} =
\bar e_{in} {\dot {\cal M} \over M} - {P \over M c^2}, 
\label{evolM}
\end{equation} 
where $M$ and $J$ are the energy-mass and the angular momentum of a
black hole, respectively; $A\equiv cJ/GM^2 = J/J_{max}$ is the
dimensionless angular momentum of the black hole; $\dot {\cal M}\equiv
\dot m \dot {\cal M}_{\rm Edd}$ is the accretion rate ($\dot {\cal
M}_{\rm Edd}$ is the Eddington accretion rate defined by $c^2 \dot
{\cal M}_{\rm Edd} \equiv L_{\rm Edd} \simeq 10^{38} M/M_\odot$ [erg
sec$^{-1}$]); $\bar e_{in}= e_{in}/c^2$ and $\bar j_{in}= c j_{in}/
GM$ are dimensionless specific energy and angular momentum of matter
at inner edge of the accretion disc; $\bar r_h = r_h c^2/GM$ is the
dimensionless black hole radius; $k = \Omega_F/\Omega_h$ (where
$\Omega_F$ is the angular velocity of the magnetic field lines
threading the horizon and $\Omega_h$ is the angular velocity of a
black hole - see Appendix A); and
\begin{equation}
P = { k(1-k) \over 32} {G^2 \over c^3} \bar r_h^2  A^2 B_{\perp}^2
M^2 \label{power} 
\end{equation} 
is the power extracted by the B-Z mechanism, where $B_{\perp}$ is
the poloidal component of the magnetic field threading the horizon.

The value of $B_{\perp}$ depends on the accretion disc model and on the
efficiency of the diffusion of the external magnetic field into a disc.
If the black hole magnetic field does not penetrate the disc but instead
interacts with it by inducing surface currents, then its energy
density can be as high as maximum total pressure in a disc, i.e.
$B_{\perp}^2 \simeq 8\pi p_{tot,max}$.

The dependence of $p_{tot,max}$ on accretion rate, viscosity and spin
for $\alpha$- and $\beta$-discs is illustrated in Figure \ref{pmax},
and respective formulas are specified in Appendix B. Using Figure
\ref{pmax} for estimations of $P$ in Eq. (\ref{power}) one should
remember that the disc pressure depends also on the black hole mass:
for radiation pressure dominated $\alpha$-discs it is $\propto 1/M$;
for radiation pressure dominated $\beta$ discs $\propto 1/M^{4/5}$;
and for gas pressure dominated discs $\propto 1/M^{9/10}$.
\begin{figure} 
\begin{center}
\leavevmode 
\epsfxsize=220pt \epsfbox{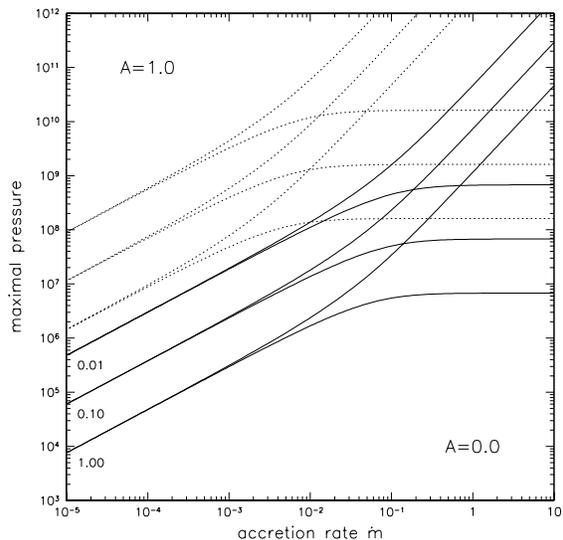} 
\end{center} 
\caption{
  Maximum pressure in the disc versus accretion rate. Dotted lines
  are for spin $A=1$ and solid lines are for $A=0$. The lines within
  a given pattern are parameterized by the value of the viscosity
  parameter.  For high accretion rates the lines split into those
  representing maximum pressure in $alpha$-discs (lower branches)
  and those representing maximum pressure in $\beta$-discs (upper
  branches). Pressure is calculated for $M_{\rm BH}=10^8 M_{\odot}$.}
\label{pmax} 
\end{figure} 
Much smaller $B_{\perp}$ are confined by a disc, if the black hole
magnetic field diffuses into a disc. Then, as was argued by Ghosh \&
Abramowicz \shortcite{ga97}, the energy density of magnetic field
which threads the black hole cannot exceed the energy density of the
magnetic field in a disc.  Assuming that the angular momentum
transport in the disc is dominated by the Maxwell stress, $t_{r\phi} =
B_r B_{\phi}/4\pi$ \cite{bh91}, one can find, using definitions of the
$\alpha$-disc ($t_{r\phi} = \alpha p_{tot}$) and of the $\beta$-disc
($t_{r\phi} = \beta p_{gas}$), that for $\alpha$-discs
$B_{\perp}^2/8\pi \sim \alpha p_{tot,max}^{(\alpha)}$, and for
$\beta$-discs $B_{\perp}^2 / 8 \pi
\sim \beta p_{gas,max}^{(\beta)}$.

For a given accretion rate, the rate of angular momentum transport is
model independent. This transport rate is given by the integral of
$t_{r\phi}$ over the disc height. In a radiation pressure dominated
accretion disc in which electron scattering is the only source of
opacity the disc semi-thickness is equal to $H=(3/4) \dot {\cal M}
r_S$, where $r_S=2GM/c^2$ is the Schwarzschild radius. Therefore,
$B_{\perp}^2 \sim t_{r\phi}$ is the same for both models and is
numerically equal to $8 \pi p_{tot, max}^{(\alpha)}$ for discs with
$\alpha=\beta=1$.The maximum pressure for such discs is $\sim 10^8
/M_8$ dyne cm$^{-2}$ which gives $P_{max} \sim 3 \times 10^{44} M_8 $
ergs s$^{-1}$, as calculated from Eq.~(\ref{power}) for $k=1/2$ and
$A=1$.  Assuming that efficiency of radio production is 10\% ($L_r
\sim 0.1 P$) and noting that most luminous radio sources have $L_r
\sim 10^{45}$ ergs s$^{-1}$ one needs $M \sim 3 \times 10^{9}
M_{\odot}$ to explain them.  The existence of such massive black holes
in some galactic nuclei is confirmed by the evaluation of the black
mass in M87 \cite{har94}.  In a model with non-diamagnetic discs it is
difficult, however, to obtain the radio-loudness of radio-loud
quasars, defined as the ratio of the radio flux to the optical flux. A
typical radio-loudness in radio-loud quasars corresponds roughly to
$P/L_d \sim 0.1$ \cite{rs91}, where $L_d = (1 - \bar e_{in}) \dot
{\cal M} c^2$ is the luminosity of an accretion disc.  Such values are
reachable by models with non-diamagnetic discs only if $L_d \le 0.01
L_{Edd}$ and it is not clear whether such objects produce UV bumps and
emission lines which are attributes of both radio-quiet and radio-loud
quasars.  Accretion flows in such systems would be rather dominated by
advection (see Lasota 1998; Narayan, Mahdevan \& Quataert 1998 for
recent reviews). The presence of such a flow in M87 was proposed by
Reynolds et al. \shortcite{rey96}.

From equations (\ref{evolA}) and
(\ref{evolM}) one gets that at equilibrium, ${\rm d}A/{\rm d}t=0$,
the radio loudness for a thin, Keplerian disc is expressed by 
\begin{equation} {P \over L_{\rm
d}}={(\bar j_{in} - 2 A \bar e_{in}) \over ( {2 \bar r_h \over  k A}
- 2 A ) (1 - \bar e_{in})} 
\end{equation} 
and is independent on the disc model chosen. The equilibrium radio
loudness as a function of the equilibrium spin is presented in
Figure~\ref{fig3}.

The evolution of the black hole spin, as derived from equations
(\ref{evolA}) and (\ref{evolM}) can be characterized by two phases.
During the first, $A$ makes its way towards the equilibrium value
$A_{eq}$ given by $dA/dt =0$, the second is described by $dA/dt \simeq
0$ and Eq. (\ref{evolM}).  Examples of such evolution for $\alpha$-
and $\beta$-discs were presented by Moderski \& Sikora (1996a; 1997).
$A_{eq}$ solutions as a function of $\dot m$ are shown in Figure 2.
For high accretion rate $\alpha$-discs these solutions are independent
of $M_{BH}$. For $\beta$-discs and gas pressured dominated discs
$A_{eq}$ depends on $M_{BH}$, but very weakly, $\propto M_{BH}^{-1/5}$
and $\propto M_{BH}^{-1/10}$, respectively.
\begin{figure}
\begin{center} 
\leavevmode 
\epsfxsize=220pt \epsfbox{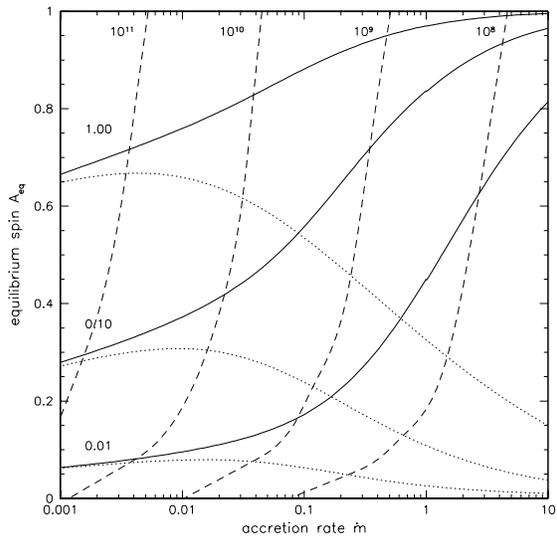}
\end{center} 
\caption{
 Equilibrium spins of black holes for $\alpha$ (solid lines)
 and $\beta$ (dotted lines) discs. Values of viscosity parameters
 are listed on the right side of the plot. Dashed lines indicate
 time scales (in years), after which the equilibrium spins are
 established. Timescales are labeled at the top of the plot. Evolutions
 were calculated for initial black hole mass $M_{\rm BH} = 10^8
 M_{\odot}$.}
\label{fig2} 
\end{figure} 
One can see in Figure~\ref{fig2}, that for $\alpha$-disc models, low
spin equilibrium solutions ($A < 0.1$ say) exist only for very low
accretion rates and only for very small $\alpha$ parameters. Such
states cannot be achieved during a Hubble time (see the dashed lines
which mark the time scales during which black holes reach $A_{eq}$)
and, therefore, black holes once spun-up to a high spin value will
rotate fast forever.  Black holes accreting from $\alpha$-discs can
have low spins only if they are formed with low spins and then accrete
very little in comparison with the mass they collected during the
formation process \cite{msl97}, or if some processes lead to random
changes of a disc angular momentum. These possibilities are discussed
in next Sections.

Black holes with low equilibrium spins for the high accretion rates
exist only if the accretion disc is a $\beta$-disc.  For the viscosity
parameter $\beta \sim 0.1$ and $\dot m \sim 10$ such models give
$A_{eq} \sim 0.05$, and if black holes start from different initial
spins they reach the equilibrium spin in $\sim 10^7$ years.  However,
as can be found from Figures 2 and 3, $P/L_d$'s for such parameters
correspond to radio-loudness of the radio-loud quasars rather than to
that of radio-quiet quasars (see Section 4). This problem could
eventually be over-passed by postulating very low efficiency of
conversion of energy extracted from a black hole to the jet. Then,
black holes with a low equilibrium spin would represent radio-quiet
quasars, while black holes reaching $A \sim 1 \gg A_{eq}$,
e.g. following coalescence of two black holes \cite{wc95}, would
represent radio-loud quasars. However, as Figure~4 shows, such black
holes spin-down in less than $10^6$ years, i.e.  in two orders shorter
time scale then typical lifetime of radio sources in quasars.
\begin{figure} 
\begin{center}
\leavevmode 
\epsfxsize=220pt \epsfbox{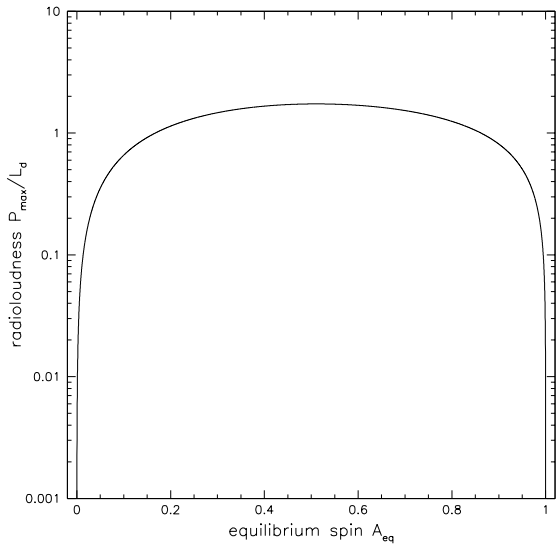} 
\end{center} 
\caption{
  $P/L_d$ vs. $ A_{eq}$ for geometrically thin accretion disc (inner
  edge of the disc on the marginally stable orbit) and for maximal
  efficiency of energy extraction ($k=1/2$).}
\label{fig3} 
\end{figure} 
\begin{figure} 
\begin{center} 
\leavevmode
\epsfxsize=220pt \epsfbox{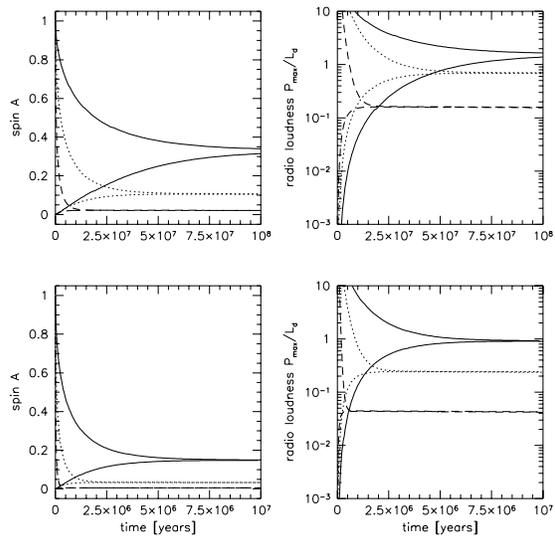} 
\end{center} 
\caption{
  Evolution of the black hole spin (left panels) and of the
  radio-loudness (right panels) for the $\beta$ accretion disc model.
  Initial values of spin are $A_0=0.0, 0.4, 1.0$. Solid lines are for
  discs with $\beta=1$; dotted lines are for $\beta=0.1$; and dashed
  lines are for $\beta=0.01$.  The upper panels are for $\dot m = 1$
  and the lower panels are for $\dot m = 10$. Evolutions calculated
  for initial black hole mass $M_{\rm BH}=10^8 M_{\odot}$. }
\label{Alow_beta} 
\end{figure}

All these considerations show that, in the framework of the spin paradigm,
it is impossible to account for the observed radio dichotomy of quasars
if the angular momentum accreted during their evolution is always of the
same sign. 
\section{SWITCHING BETWEEN  PRO- AND RETROGRADE ACCRETION DISCS} 
\begin{figure} 
\begin{center} 
\leavevmode
\epsfxsize=220pt \epsfbox{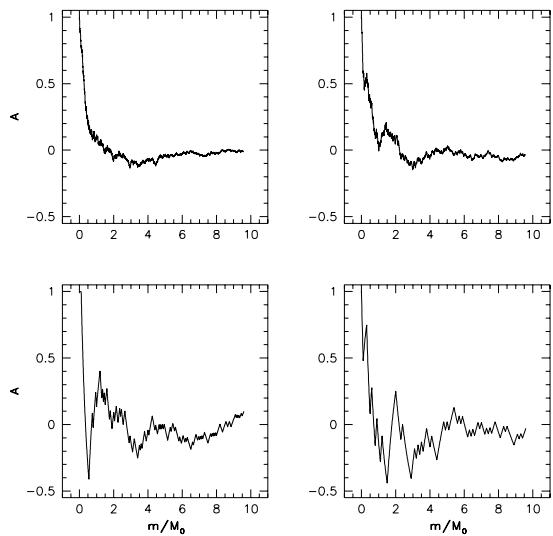} 
\end{center} 
\caption{
  Examples of spin evolution for different masses accreted each
  time. All evolutions are from initial spin $A_0 = 1$. From the upper
  left to bottom right we have $\Delta m = 0.005, ~0.01, ~0.05, ~0.1$,
  where $\Delta m$ is the ratio of the mass of the accreted portion
  to the initial mass of the black hole $M_0$.}
\label{examples} 
\end{figure}

\begin{figure} 
\begin{center}
\leavevmode 
\epsfxsize=220pt \epsfbox{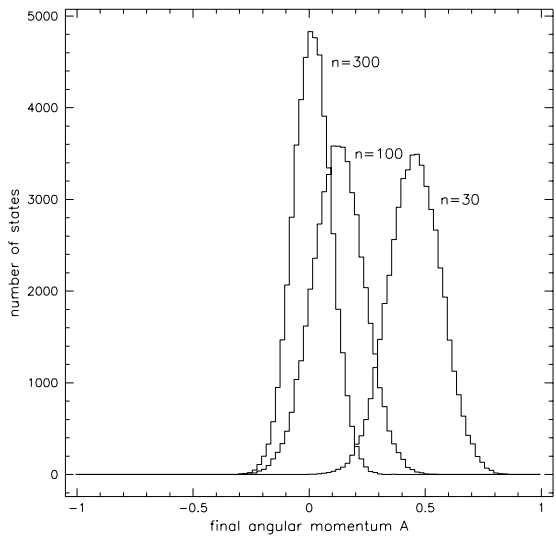} 
\end{center} 
\caption{
  Histogram of the final states of 50000 evolutionary tracks computed
  for $A_0=1$ and $\Delta m=0.01$. The number of accretion events is
  marked on the curves.}
\label{hist} 
\end{figure}

\begin{figure} \begin{center}
\leavevmode 
\epsfxsize=220pt \epsfbox{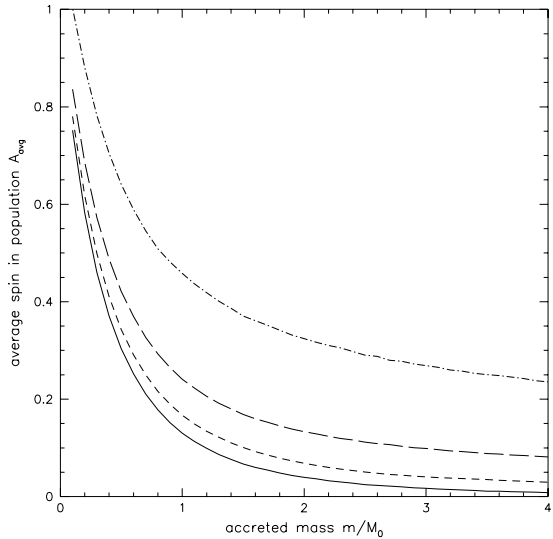} 
\end{center} 
\caption{
  Average spin and spread in the population of black hole evolving
  from $A_0=1$ as a function of accreted mass. Average spin is marked
  with solid line while $1\sigma$ standard deviation in distribution
  (assuming Gaussian shape) is marked as dotted, dashed and long
  dashed lines for $\Delta m=0.001, 0.01$ and $0.1$ respectively. Only
  $+1\sigma$ contours are plotted}
\label{retfinal} 
\end{figure}

A spinning black hole can be decelerated if it accretes angular
momentum of the opposite sign.  Since a simple version of retrograde
accretion was discussed in details in one of our previous papers
\cite{ms96b}, here we present only results which are directly
connected to the possibility of spinning down of black holes.  As was
shown by Moderski and Sikora \shortcite{ms96b}, it suffices to accrete
$\sim 0.2$ of the initial black hole mass to decelerate a black hole
from its maximum spin to zero. However, accretion of an larger amount
larger than $0.2$ of the initial mass will spin-up a black hole again,
so that to maintain a low value of the time-averaged spin, the
evolution of a black hole must be governed by many ``small'' accretion
events with random orientation of the accreted angular momentum. In
the calculations we assume that portions (supplied e.g.  by giant
molecular clouds or by the capture of dwarf galaxies) all have equal
masses. Because of the Bardeen-Petterson effect \cite{bp75} the disc,
near the black hole, rotates in its equatorial plane we randomly
choose only between direct or retrograde accretion and we did not
consider more complicated alignments. Some examples of the evolution
of the black hole spin for several values of the mass accreted at each
accretion event are presented in Figure~\ref{examples}.
We can see from Figure~\ref{examples} that spinning down of a black
hole with alternating direct--retrograde discs is very efficient and
leads to black holes with spin values fluctuating around zero. To
check whether this effect depends on the mass of the accreted portion
of matter we performed a number of numerical evolutions for different
ratios of the accreted mass to the initial mass of the black hole,
$\Delta m$. We traced the evolution of 50 thousands systems evolving
from $A=1$ until accreted mass exceeded four times the initial mass
of the black hole and studied how the black hole spin changed with
the number of accretion events.  An example of spin distributions
in the simulated population for three different numbers of accretion
episodes and for $\Delta m=0.01$ is presented in Figure~\ref{hist}.
We found that value of the peak does not depend on the mass $\Delta m$
but only on total accreted mass. and that the spread in distribution is
wider for higher $\Delta m$.  In Figure~\ref{retfinal} we show the mean
spin value and the spread in the population as a function of accreted
mass.  It is worth noting that results do not change significantly
if we consider geometrically thick discs, in which the inner disc
radius can be close to the marginally bound orbit \cite{al80}. We
checked that  masses smaller by only 10\% are needed to establish the
same population of black holes as in the case of thin discs.

One should mention that results presented in this Section were obtained
assuming that $P=0$. With this assumption, the black hole evolution 
is described by a simple equation 
\begin{equation} 
{dA \over d \ln M}
=  {\bar j_{in} \over \bar e_{in}} - 2 A , 
\label{Pzero} 
\end{equation}
which for a disc edge located at the marginally stable ($r_{ms}$) or
marginally bound ($r_{mb}$) orbit has an analytical solution (see
Bardeen 1970 and Moderski \& Sikora 1996b).  Our results can be used
to describe evolution of the black hole also for non-zero $P$,
provided we follow only these parts of evolutionary tracks for which
$A < A_{eq}$, because at such $A$'s the evolution is dominated by
accretion processes, rather than by the B-Z mechanism and, therefore,
the results are insensitive to the value of $P$.
\section{DISCUSSION}
Radio dichotomy of quasars was discovered many years ago (Strittmatter
et al. 1980; Kellerman et al. 1989), but is still waiting for a
theoretical explanation.  As for now, the consensus concerns only one
aspect of the problem: it is clear that jets in quasars must be formed
near a supermassive black hole.  This follows from the energetics of
quasar jets, since no other known sources could power jets at a rate
reaching $10^{46}$ ergs s$^{-1}$ for millions of years (Rawlings \&
Saunders 1991; Leahy et al. 1989). Independent argument for the
formation of extra galactic jets in the vicinity of supermassive black
holes is provided by direct VLBI observations of nearby radio
galaxies.  In particular, in 3C 274 (M87) the jet is seen down to
$10^{16}$ cm from the center \cite{jb95}, which corresponds to $100$
gravitational radii for the $3 \times 10^9 M_{\odot}$ \cite{har94}
central black hole.  There, very deep in the gravitational potential
well, jets could be powered either by the innermost parts of an
accretion disc (Blandford \& Payne 1982; Park \& Vishniac 1994;
Contopoulos 1995; Begelman 1995) or by a rotating black hole
(Blandford \& Znajek 1977; Rees et al. 1982).  However, no jet
production model can be successful, if it fails to explain why only a
small fraction of quasars is radio loud and why radio-loudness has a
bimodal distribution.

In terms of the $R=F_r/F_o$ ratio, where $F_r$ and $F_o$ are the
monochromatic fluxes measured at frequencies $\sim 10^{10}$ Hz and
$\sim 10^{15}$ Hz, respectively, radio-quiet quasars cluster around $R
\sim 0.3$ and radio-loud quasars cluster around $R \sim 300$
(Kellermann et al. 1989; Falcke, Sherwood \& Patnaik 1996).  Thus, the
average radio-loudness of the two quasar populations differs by a
factor $10^3$ and this number, together with typical radio
luminosities of radio loud quasars, $L_r \sim 10^{45}$ ergs s$^{-1}$,
provides the basic quantitative conditions which should be satisfied
by any unified model of quasars.  These conditions, together with our
results discussed in the two previous sections are used below to test
spin based models of a jet activity in quasars. The predictions of
such models should also satisfy such observationally established
trends, as

\noindent - radio-loud quasars avoid  disc-galaxies and have
UV-luminosities $ \ge 10^{46}$ ergs s$^{-1}$;

\noindent - radio-quiet quasars are present both in spiral and
elliptical galaxies \cite{tay96} and their radio properties do not
depend on the galaxy morphology \cite{kuk98};

\noindent - radio properties of radio-quiet quasars suggest that they
are, like in radio loud quasars,  related to the jet production by
a central engine.

Assuming that the efficiency of conversion of jet energy into radio
emission is $\sim 10$\%, the typical jet in radio-loud quasars should
have $P~\sim~10^{46}$ ergs s$^{-1}$. Similar jet powers are deduced by
calculating the total energy content of extended radio sources and
dividing it by the age of the source \cite{lea89}, or from energetics
of $\gamma$-ray production in sub-parsec jets (see, e.g., Sikora
1997).  Largest powers which can be extracted from rotating black
holes are given by equation (\ref{power}). For $A=1$ and $B_{\perp} =
8\pi p_{tot}$ we obtain $P_{max} \sim 3 \times 10^{44} M_8^2
p_{tot,8}$ ergs s$^{-1}$, where $p_{tot,8} = p_{tot}/10^8$dyne
cm$^{-2}$ and $M_8 = M/10^8 M_{\odot}$.  One can see from
Figure~\ref{pmax} that for high accretion rates ($\dot m > 0.1$, say)
and black hole masses $\sim M=10^9 M_{\odot}$, a pressure $\ge 10^8$
dyne cm$^{-2}$ is provided by $\alpha$-discs with $\alpha \le 0.1$,
and by all $\beta$-discs.  Thus, the B-Z mechanism is efficient enough
to power jets in radio-loud quasars, provided the black hole magnetic
field is supported by the total disc pressure.  If the latter is not
true and, as Ghosh and Abramowicz \shortcite{ga97} argued, the energy
density of the black hole magnetic field cannot exceed the energy
density of the maximum magnetic field in a disc, then the maximum
pressure of the black hole's magnetic field is numerically equal to
the total pressure in an $\alpha=1$ disc. In this case black hole
masses $\sim 3 \times 10^9 M_{\odot}$ are required in order to get $P
\sim 10^{46}$ ergs s$^{-1}$. The case of M87 seems to prove that such
black holes are not necessarily exceptional \cite{har94}.  However,
one should note here, that the question of the diffusion of an
external magnetic field into an accretion disc is still open (see,
e.g., Wang 1995; Bardou \& Heyvaerts 1996).

Assuming, as before, that the fraction of the jet energy converted into
radiation is 10\%, and that the bolometric corrections for jet radiation at
$\sim 10^{10}$Hz and for accretion disc radiation at $\sim 10^{15}$Hz
are of the same order, we obtain that $P/L_d \sim 10 L_r/L_o \sim 10
F_r{\nu}_r/F_o{\nu}_o \sim 10^{-4} R$. Thus, for radio-loud quasars
models should predict $P/L_d \sim 0.1$, while radio-quiet quasars
should cluster around $P/L_d \sim 10^{-4}$. As is seen from
Figures~2 and 3, $P/L_d \sim 0.1$ nicely corresponds to black hole
equilibrium spin solutions for all but $\alpha > 0.1$ disc models. One
can also check, that there are no equilibrium spin solutions which
would correspond to radio-loudness of radio-quiet objects. For them
$A < 0.03$ is required, provided that radio luminosity scales linearly
with $P$. A population of such low spin black holes can exist only if
black holes are born with very low spin and then accrete very little
(Moderski, Sikora \& Lasota 1997), or if black hole evolution is
determined by multi-accretion events with random angular momenta.

As one can deduce from results presented in Figure~7, hundreds of
accretion events per object are required in order to have more than
$90$ \% of black holes with $A < 0.1$ at any given moment. This is too
much to be obtained by accretion events induced by capture of dwarf
galaxies, but can be achieved by accretion of molecular clouds.
Molecular cloud accretion events were recently proposed by Sanders
\shortcite{san98} to explain some properties of Sgr A$^*$ and other
AGNs.  This scenario is supported by the random orientation of central
engines vs. the orientation of galactic discs, as deduced from
observations of ``UV'' cones (Wilson \& Tsvetanov 1994; McLeod \&
Rieke 1995) and radio axis \cite{ckp98} in Seyfert galaxies. Here we
should note, that in our simplified treatment of the multi-accretion
scenario (Section~3), we didn't take into account the coupling between
the spin of the black hole and the orbital angular momentum of the
approaching molecular clouds. Such coupling supposedly leads to random
wondering of the black hole spin vector.

One can now speculate that changes of orientation of the black hole
spin could be interrupted and the black hole could be spun-up to very
high spins following a merger process. This process could induce a
massive and long lasting accretion event. If during such an event the
accretion proceeds from a fixed plane and at least doubles the black
hole mass, the black hole spin reaches the equilibrium spin and the
object becomes a typical radio-loud quasar \cite{msl97}.

Since mergers happen mostly in groups and clusters of galaxies, where
the population of galaxies is dominated by ellipticals, this could
explain why radio loud quasars avoid spiral galaxies. Observational
arguments for such a scenario are exactly the same as those used
by Wilson and Colbert \shortcite{wc95}. The only difference is, that they
postulated formation of high spin black holes via coalescence of two
supermassive black holes, while in our scenario high spins result from
an accretion process. Note, however, that a coalescence of two black
holes, if it happens, does not have to affect much our scenario.
If the coalescence involves two black with very different masses,
the final spin will be determined by the accretion process, otherwise
both processes lead to similar spins.

What are the perspectives for an observational test of the assumption
that radio-quiet objects have low spins?  A possibility to measure the
spin of supermassive black holes is provided by the detailed studies
of profiles of the X-ray fluorescent iron line produced in the surface
layer of the innermost parts of accretion discs. Such lines are
detected in many Seyfert galaxies, which represent the low luminosity
branch of radio-quiet quasars.  For at least one of such objects the
line profile was claimed to be consistent with the kinematics given by
the rotation of a disc around a black hole in fast rotation
\cite{iwa96}. However, as demonstrated by Reynolds and Begelman
\shortcite{rb97}, similar line profiles can be produced around
non-rotating black holes, provided that a large part of the line
emission comes from below the marginally stable orbit.  Therefore,
much more detailed theoretical models and sensitive observations are
required to get conclusive diagnostics from this type of
investigations.

The remarkable discovery of relativistic jets in several Galactic
X-ray sources (cf. Mirabel \& Rodriguez 1994; Hjellming \& Rupen 1995;
Newell, Spencer \& Garrett 1997) suggests that the radio-dichotomy
exists for Galactic compact objects as well. As was argued recently by
Zhang, Cui \& Chen \shortcite{zcc97}, jet activity in these sources
can also be conditioned by the value of the black hole spin.
\section*{ACKNOWLEDGMENTS} 
RM and MS acknowledge support from KBN grant 2P03D01209. During his
stay in Meudon RM was supported by the R\'eseau Formation Recherche of
the French Minist\`ere de l'Enseignement Sup\'erieure et de la
Recherche, he also acknowledges the support from Foundation for Polish
Science Fellowship. JPL thanks the Isaac Newton Institute for
hospitality in April 1998.

\appendix

\section{DIMENSIONLESS QUANTITIES IN THE KERR METRIC}

The black hole radius in the Kerr metric is 
\begin{equation}
\tilde r_{\rm h}= c^2 r_{\rm h}/GM = 1 + (1 - A^2)^{1/2}.  
\label{rh}
\end{equation} 
The black hole angular velocity is 
\begin{equation}
\tilde \Omega_{\rm h} = {G M \over c^3} \Omega_{\rm h} = {A \over 2
\tilde r_{\rm h}}.  
\label{omegah} 
\end{equation} 
The marginally stable orbit is (upper signs are for direct accretion
and lower sign for retrograde accretion)
\begin{equation} 
\begin{array}{rcl} 
\tilde
r_{\rm ms} &=& c^2 r_{\rm ms}/GM = \\ &=&3+ Z_2 - \mp \left[ (3-Z_1)
(3+Z_1+2Z_2) \right]^{1/2} ~ , 
\end{array} 
\label{rms} 
\end{equation}
where 
\begin{equation} Z_1=1+ (1-A^2)^{1/3} \left[ (1+A)^{1/3} +
(1-A)^{1/3} \right] 
\end{equation} 
and 
\begin{equation} Z_2= (3
A^2 + Z_1^2)^{1/2}.  
\end{equation} 
and inversely: 
\begin{equation}
A=\pm {1 \over 3} \tilde r_{\rm ms}^{1/ 2} \left( 4-{(3 \tilde r_{\rm
ms}-2)}^{1/2} \right).  
\label{Ams} 
\end{equation} 
The specific energy and specific angular momentum of a particle on the
marginally stable orbit are, respectively,
\begin{equation} \tilde e_{\rm
ms}=  e_{\rm ms}/c^2 = {\left( 1 - {2 \over {3 \tilde r_{\rm ms}}}
\right)}^{1/2} 
\label{ems} 
\end{equation} 
and 
\begin{equation}
\tilde j_{\rm ms} = {cj_{\rm ms} \over GM} = \pm {2\over 3\sqrt{3}}
\left[ 1+2{(3\tilde r_{\rm ms}-2)}^{1/2} \right]  .  
\label{jms}
\end{equation} 
\section{PRESSURE IN $\alpha$- AND $\beta$-DISCS}
For high accretion rates pressure in geometrically thin discs around
supermassive black holes is dominated by radiation and can be
approximated by formulas (Novikov \& Thorne 1973; Sakimoto \& Coroniti
1981)
$$ p_{rad}^{(\alpha)} = 3.0 \times 10^8 \alpha^{-1} M_8^{-1}
\bar r^{-3/2} {{\cal B}^2 {\cal E} \over {\cal A}^2} 
\eqno (B1) 
$$ 
for $\alpha$-discs and 
$$ p_{rad}^{(\beta)} = 1.6 \times 10^{14}
\dot m^{8/5} \beta^{-4/5} M_8^{-4/5} \bar r^{-18/5} {{\cal Z}^{8/5}
\over {\cal B}^{8/5} {\cal D}^{4/5}} 
\eqno (B2) 
$$ 
for $\beta$-discs, where $\dot m= \dot {\cal M} c^2 / L_{Edd}$, and
${\cal A}$, ${\cal B}$, ${\cal D}$, ${\cal E}$, ${\cal Z}$ are
relativistic corrections and can be found in Novikov \& Thorne (1973).

For low accretion rates the pressure is dominated by a gas and is 
$$ 
p_{gas}= 2.5 \times 10^{11} \beta^{-9/10} \dot m^{4/5} M_9^{-9/10}
\tilde r^{-51/20} {{\cal B}^{1/5}{\cal E}^{1/2}{\cal Z}^{4/5}\over
{\cal A}{\cal D}^{2/5}} . 
\eqno (B3) 
$$ 
\end{document}